\documentclass[12pt]{article}

\usepackage{algorithmic,amsmath,amssymb,graphics,amsthm}
\usepackage{defs}

\begin{document}
\newcommand{\dualG}{\mat{G}^{\perp}}
\newcommand{\dualcbook}{\cbook^{\perp}}
\newcommand{\event}{\mathcal{E}}

\renewcommand{\textfraction}{0}

\title{Iterative Quantization Using Codes On Graphs}
\author{\normalsize
Emin Martinian \\
[-5pt] \small Massachusetts Institute of Technology  \\
[-5pt] \small Cambridge, MA 02139\\
[-5pt] \small emin@allegro.mit.edu\\
\and 
\normalsize Jonathan S. Yedidia \\
[-5pt] \small Mitsubishi Electronic Research Labs \\
[-5pt] \small Cambridge, MA 02139\\
[-5pt] \small yedidia@merl.com
}

\date{}

\maketitle

\thispagestyle{myheadings}
\markboth{Proceedings of the 41st Annual Allerton Conference, (Monticello, IL)
  2003}{Proceedings of the 41st Annual Allerton Conference, (Monticello, IL)
  2003}
\pagenumbering{gobble}
\pagestyle{plain}

\begin{abstract}
We study codes on graphs combined with an iterative message passing
algorithm for quantization.  Specifically, we consider the binary erasure
quantization (BEQ) problem which is the dual of the binary erasure
channel (BEC) coding problem.  We show that duals of capacity
achieving codes for the BEC yield codes which approach the minimum
possible rate for the BEQ.  In contrast, low density parity check codes
cannot achieve the minimum rate unless their density grows at least 
logarithmically with block length.  Furthermore, we show that duals
of efficient iterative decoding algorithms for the BEC yield efficient
encoding algorithms for the BEQ.  Hence our results suggest that
graphical models may yield near optimal codes in source coding as well
as in channel coding and that duality plays a key role in such constructions.
\end{abstract}

\section{Introduction}

Researchers have discovered that error correction codes defined on
sparse graphs can be iteratively decoded with low complexity and
vanishing error probability at rates close to the Shannon limit.
Based on the close parallels between error correction and data
compression, we believe that similar graphical codes can approach the
fundamental limits of data compression with reasonable complexity.
Unfortunately, the existing suboptimal channel decoding algorithms for
graphical codes generally fail unless the decoder input is already
near a codeword.  Since this is usually not the case in source
coding, either a new type of graph or a new suboptimal algorithm (or
both) is required.

Before developing iterative quantization techniques it is worth
investigating the potential gains of such an approach over existing
systems.  For asymptotically high rates, when compressing a
continuous source with finite moments relative to mean square error
(MSE) distortion, entropy coded scalar quantization (ECSQ) is 1.53 dB
from the rate-distortion limit \cite{Gray_1998}.  For moderate rates
the gap is larger: in quantizing a Gaussian source relative to MSE
distortion, ECSQ systems are 1.6--3.4 dB away from the rate-distortion
limit.  For these parameters, trellis coded quantization (TCQ) using a
256-state code with optimal quantization has a gap of 0.5--1.4 dB
\cite{Marcellin_1990}.  For higher rates, sources with larger tails
(\eg, a source with a Laplacian distribution), or sources with memory,
the gaps are larger.  Thus for memoryless sources quantized at
moderate rates, new codes have the potential to improve performance by
the noticeable margin of a few decibels.  More generally, the codes on
graphs paradigm may prove valuable in a variety of scenarios involving
speech, audio, video and other complicated sources.

To illustrate possible approaches to developing graphical codes we
focus on the binary erasure quantization (BEQ) problem which is the
dual of the binary erasure channel (BEC) coding problem.  First we
describe the BEQ problem model in \secref{sec:quantization-model}.
Next, in \secref{sec:codes-eras-quant} we present our main result for
the BEQ: duals of low density parity check codes can be analyzed,
encoded, and decoded by dualizing the corresponding techniques for the
BEC.  Specifically, by dualizing capacity achieving codes for the BEC
we obtain rate-distortion approaching codes for the BEQ.  Finally, we
close with some concluding remarks in \secref{sec:concluding-remarks}.

\pagenumbering{arabic}
\addtocounter{page}{1}

\section{Quantization Model}
\label{sec:quantization-model}

Vectors and sequences are denoted with an arrow (\eg, $\genericS{x}$).
Random variables or random vectors are denoted using the sans serif
font (\eg, $\genericRV{x}$ 
or $\genericRVS{x}$).  We consider the standard (memoryless) data compression
problem and represent an instance of the problem with the tuple 
$(\srcAlph,p_{\rvSrc}(\src),\dist{\cdot,\cdot})$ 
where $\srcAlph$ represents the source alphabet, $p_{\rvSrc}(\src)$
represents the source distribution, and $\dist{\cdot,\cdot}$
represents a distortion measure.  Specifically, a source $\rvsSrc$
consists of a sequence of $n$ random variables $\rvSrcC{1}$,
$\rvSrcC{2}$, $\ldots$, $\rvSrcC{n}$ each taking values in $\srcAlph$
and generated according to the distribution $p_{\rvsSrc}(\sSrc) =
\prod_{i=1}^n p_{\rvSrc}(\srcC{i})$.
A rate $\rate$ encoder $\encoder{\cdot}$ maps $\rvsSrc$ to an integer
in $\{1, 2, \ldots, 2^{n\rate}\}$, and the corresponding decoder
$\decoder{\cdot}$ maps the resulting integer into a reconstruction
$\rvsRecon$.  Distortion between the source $\rvsSrc$ and the
reconstruction $\rvsRecon$ is measured via $D =
\frac{1}{n}\sum_{i=1}^n \dist{\rvSrcC{i},\rvReconC{i}}$. 

Shannon derived the minimum possible rate required by any data compression
system operating with distortion $D$.  The so-called rate-distortion
function is given by the formula
\begin{equation}
\label{eq:rd-formula}
R(D) =
\min_{p_{\rvRecon|\rvSrc}(\recon|\src):E[\dist{\rvSrc,\rvRecon}]
  \leq D} I(\rvSrc;\rvRecon)
\end{equation}
where $I(\cdot;\cdot)$ denotes mutual information and $E[\cdot]$
denotes expectation.

\subsection{Binary Erasure Quantization}

To highlight connections between error correction and data compression,
we consider the binary erasure quantization (BEQ) problem where the
source vector consists of ones, zeros, and ``erasures'' represented by
the symbol $\erasure$.  Neither ones nor zeros may be changed, but
erasures may be quantized to either zero or one.  Practically,
erasures may represent source samples which are missing, irrelevant,
or corrupted by noise and so do not affect the distortion regardless
of the value they are assigned.  Formally, the BEQ problem with
erasure probability $e$ corresponds to
\begin{subequations}
\begin{align}
\srcAlph &= \{0,1,\erasure\}\\
p_{\rvSrc}(\src) &= \frac{1-e}{2} \cdot\delta(\src) + \frac{1-e}{2}
\cdot\delta(\src-1) + e \cdot\delta(\src-\erasure)\\
\dist{a,b} &= 
0 \textnormal{ if } a = \erasure \textnormal{ or } a = b,
\textnormal{ and } 1 \textnormal{ otherwise.}
\end{align}
\end{subequations}
It is straightforward to show that for $D=0$ the distribution 
\begin{equation}
p_{\rvRecon|\rvSrc}(\recon|\src) = 
\delta(\recon-\src) \textnormal{ if } \src \in \{0,1\}, 
\textnormal{ and }
\frac{1}{2} \cdot\delta(\recon) + \frac{1}{2}
\cdot\delta(\recon-1) \textnormal{ if } \src = \erasure.
\end{equation}
optimizes \eqref{eq:rd-formula} and yields the value of the rate-distortion
function at $D=0$:
\begin{equation}
\label{eq:bec-rd-formula}
R_{\mathrm{BEQ}}(D=0) = 1-e.
\end{equation}

\section{Codes For Erasure Quantization}
\label{sec:codes-eras-quant}

It is well-known that the encoder for a quantizer serves a
similar function to the decoder for an error correcting code in the
sense that both take a vector input (\ie, a source to quantize or
channel output to decode) and map the result to bits (\ie, the
compressed source or the transmitted message).  The
decoder for a quantizer can similarly be identified with the encoder for an
error correcting code in the sense that both take bits as input and
produce a vector (\ie, a source reconstruction or a channel
input).  Thus it is natural to investigate whether swapping the
encoder and decoder for a good error correcting code such as a
low density parity check (LDPC) code produces a good quantizer.

\subsection{LDPC Codes Are Bad Quantizers}

\begin{figure}[hbpt]
\begin{center}
\resizebox{2.75in}{!}{\includegraphics{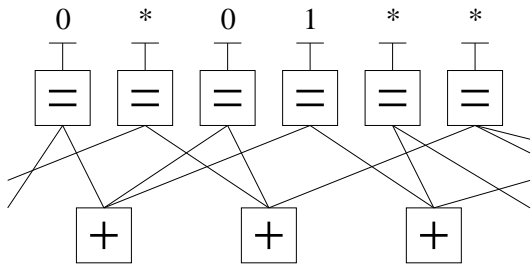}}
\caption{Using an LDPC code for binary erasure quantization.  The
  boxes with $=$ signs are repetition nodes: all edges connected to an
  $=$ box must have the same value.  The boxes with $+$ signs are
  check nodes: the modulo 2 sum of the values on edges connected to a
  $+$ box must be 0.  The source consists of 0's, 1's, and erasures
  represented by $*$'s.  Erasures may be quantized to 0 or 1 while
  incurring no distortion.  A non-zero distortion must be incurred for
  the source shown above since the left-most check cannot be
  satisfied.
  \label{fig:beq_ldpc}}
\end{center}
\end{figure}

One benefit of studying the BEQ problem is that it demonstrates why
low density parity check (LDPC) codes are inherently unsuitable for
quantization.  Specifically, consider an LDPC code like the one
illustrated in \figref{fig:beq_ldpc} using Forney's normal graph
notation \cite{it:2001:forney}.  If all the variables connected to a
given check are not erased, then there is an even chance that no code
symbol can match the source in that position and thus the distortion will be
positive regardless of the code rate.  Thus, as stated in
\thrmref{th:ldpc-codes-bad} and proved in \appref{app:proofs},
successful decoding is asymptotically unlikely unless the density of
{\em every} parity check matrix for the code increases logarithmically
with the block length.%
\footnote{We may expect the density of a code to
increase as it approaches the {\em capacity} or {\em rate-distortion
function}, but a code whose density also increases with 
{\em block length} seems undesirable.} 

\begin{thrm}
\label{th:ldpc-codes-bad}
Let $\cbook_{(n)}$ be a sequence of linear codes of length $n$ and
fixed rate $R$ such that the probability that binary erasure
quantization using $\cbook_{(n)}$ of a random source sequence with
$e \cdot n$
erasures will succeed with zero distortion is bounded away from 0 as
$n \rightarrow \infty$.  Then regardless of the values of $R$ and $e$,
the degree of the parity-check nodes in any parity-check graph
representation of $\cbook_{(n)}$ must increase at least
logarithmically with $n$.
\end{thrm}

\vspace{-.175in}

The poor performance of LDPC codes for quantization may seem
surprising in light of their excellent properties in channel coding,
but it has long been recognized that good codes for error correction
and quantization may be different.  The former is essentially
a packing problem where the goal is to place as many codewords as
possible in a given space such that the codewords are far apart and
can be distinguished despite noise.  The latter is a covering problem
where the goal is to place as few codewords as possible in a given
space such that every point in space is near at least one codeword.

From any good error correcting code (respectively data compression
code) it is easy to obtain another code which is almost as good at
error correction (resp. data compression) but terrible at source
coding (resp. channel coding).  For example, removing half the
codewords has an asymptotically negligible effect on error correction
since it only decreases the rate by $1/n$ and only increases
robustness.  But, removing half the codewords can
dramatically hinder source coding since half the time the source may
be very far from the nearest codeword.  Conversely, doubling the
number of codewords has an asymptotically negligible effect on data
compression since the rate only increases by $1/n$ while the
distortion may decrease slightly.  But doubling the number of
codewords can be catastrophic for error correction if it drastically
reduces the distance between codewords.

\subsection{Dual LDPC Codes}

Many researchers have explored duality relationships between error
correction and source coding.  Such work demonstrates that
often a good solution for one problem can be obtained by dualizing a
good solution to the other. 
%
%
Continuing in this tradition, we study the properties of dual LDPC
codes for binary erasure quantization.

Formally, a length $n$ binary linear code $\cbook$ is a subspace of
the $n$ dimensional vector space over the binary field and the dual
code $\dualcbook$ is the subspace orthogonal to $\cbook$.  For
LDPC codes, the code $\cbook$ is usually specified by the parity check
matrix $\mat{H}$ representing the constraint that $\mat{H}\genericS{x}^{T} =
0$ if and only if $\genericS{x}$ is a codeword.  To obtain the dual code
$\dualcbook$ we can recall that the generator $\dualG$ of
$\dualcbook$ is exactly $\mat{H}$.  If the
code $\cbook$ is represented by a normal graph as in
\figref{fig:beq_ldpc}, then the graph of the dual code $\cbook$ can be
obtained by swapping $+$ and $=$ nodes
\cite{it:2001:forney}.  In dualizing the code graph in this
manner it may be useful to note that while the graph of
$\cbook$ obtained from $\mat{H}$ represents a syndrome former for
$\cbook$, the dualized graph represents an encoder for $\dualcbook$.  

\begin{figure}[hbtp]
\begin{center}
\resizebox{2.75in}{!}{\includegraphics{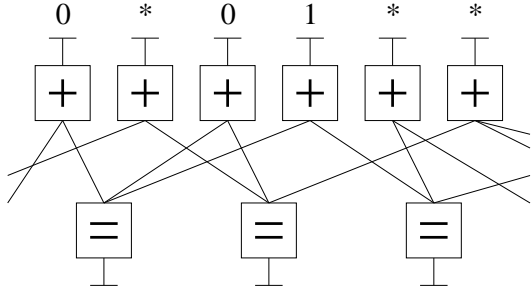}}
\caption{Using the dual of an LDPC code for binary erasure
  quantization.  Choosing values for the
  variables at the bottom produces a codeword.  The values for each
  sample of the resulting codeword are obtained by taking the sum
  modulo 2 of the connected variables.  In contrast to
  \figref{fig:beq_ldpc}, this structure can successfully match the
  source with no distortion if the bottom 3 variables are set to 0, 0,
  1.  \label{fig:beq_dual_ldpc}}
\end{center}
\end{figure}

For example, \figref{fig:beq_dual_ldpc} is obtained by dualizing the
code graph in \figref{fig:beq_ldpc}.  Notice that while the original
code cannot quantize the source with distortion 0, the dual code can.
Intuitively, the advantage of a low density encoder structure is that
it provides a simple representation of a basis which can be used to
construct the desired vector.  In the following sections, we
investigate the properties of dual LDPC codes for quantization with
both optimal quantization and iterative quantization.

\vspace{-.175in}

\subsection{Optimal Quantization/Decoding and Duality}

The following theorem (proved in \appref{app:proofs}) demonstrates the
dual relationship between channel decoding and source quantization
using optimal decoding/quantization algorithms.

\begin{thrm}
\label{thrm:dual-optim-quant}
A channel decoder for the code $\cbook$ can correctly decode every
received sequence with the erasure pattern%
\footnote{If symbol $i$ is erased (resp. unerased) then $e_i = 1$
 (resp. $e_i = 0)$ in our notation for erasure patterns.}
$\genericS{e}$ if and only if a quantizer for the code $\dualcbook$ can
successfully quantize every%
\footnote{Note that some source sequences (\eg, the all zero sequence)
  can be successfully quantized using $\dualcbook$ regardless of
  $\genericS{e}^{\perp}$.  Similarly, a system which decodes ambiguous
  received sequences to the all zero sequence may succeed even when
  many erasures occur.  Thus to obtain the desired equivalence between
  correct decoding and successful quantization we define correct
  decoding (resp. successful quantization) as being able to deduce the
  transmitted codeword (resp. a codeword matching non-erased positions
  of the source) for every possible received sequence (resp. source)
  with the the erasure pattern $\genericS{e}$
  (resp. $\genericS{e}^{\perp}$).}
source sequence with the erasure pattern $\genericS{e}^{\perp} = 1 -
\genericS{e}$. 
\end{thrm}

From this result we immediately obtain the following Corollary.
\begin{corol}
Let $\cbook_{(n)}$ be a sequence of linear codes which achieves the
capacity of a binary erasure channel with erasure probability $e$
using optimal decoding.  The sequence $\dualcbook_{(n)}$
obtained by taking the duals of $\cbook_{(n)}$
achieves the minimum rate for $D=0$ for the BEQ with erasure
probability $e^{\perp} = 1-e$ using optimal quantization.
\end{corol}

The statement and proof of the two preceding results contain
a curious duality between erased/known symbols in source coding and
known/erased symbols in channel coding.  A similar duality exists
between a likelihood ratio, $\lambda$, and its Fourier
transform $\Lambda = \frac{1 - \lambda}{1 +
  \lambda}$ used in dualizing the
sum-product algorithm \cite[pp. 545--546]{it:2001:forney}.
Specifically, the Fourier transform maps known/erased likelihood
ratios to erased/known likelihood ratios.

\subsection{Iterative Decoding/Quantization and Duality}

In the following we first review the intuition behind iterative
erasure decoding algorithms and describe the
particular decoding algorithm we consider in
\tabref{tab:erasure-dec-alg}.  Next we outline the intuition behind a
similar approach for iterative quantization and precisely describe our
quantization algorithm in \tabref{tab:erasure-quant-alg}.  Finally, we
show that these algorithms are duals.

\subsubsection{Iterative Erasure Decoding}

Many iterative message-passing decoding algorithms are 
essentially based on the following idea.  The
outgoing message on edge $i$ of a + node is the modulo-2 sum of all
incoming messages (excluding edge $i$) with the proviso that if any
incoming message (excluding edge $i$) is $*$ then the outgoing message
is also $*$.  For an = node, the outgoing message on edge $i$ is $*$
only if all other incoming messages are $*$, otherwise the outgoing
message is the same as the known incoming message or messages.  These
message-passing rules can be interpreted as determining the outgoing
message on edge $i$ by applying the following 
``sum'' and ``product'' formulas to all other incoming messages.
\footnote{In the product rule for erasure decoding, the symbol \#
  denotes a contradiction which is impossible if only erasures and no
  errors occurred.} 

\begin{scriptsize}
\begin{equation}
\label{eq:bec-rules}
\begin{array}{c||c|c|c}
+ & 0 & 1 & * \\ \hline\hline
0 & 0 & 1 & * \\ \hline
1 & 1 & 0 & * \\ \hline
* & * & * & * \\ 
\end{array}
\textnormal{\normalsize{ \ \ and \ \ }}
\begin{array}{c||c|c|c}
\times & 0 & 1 & * \\ \hline\hline
0      & 0 & \textnormal{\#}  & 0 \\ \hline
1      & \textnormal{\#} & 1 & 1  \\ \hline
*      & 0 & 1 & *  \\
\end{array}
\end{equation}
\end{scriptsize}

It is well-known that such algorithms yield optimal
decoding on a tree and also perform well on graphs with cycles provided 
appropriate scheduling and initialization rules are selected.
Initializing all messages to $*$ and using sequential or 
parallel schedules are common choices.  For the purpose of proving
theorems, we consider a sequential schedule in the
\alg{ERASURE-DECODE} algorithm of \tabref{tab:erasure-dec-alg}.

\vspace{-.1in}
\begin{table}[htb]
\caption{An algorithm for iteratively decoding data with
  erasures. \label{tab:erasure-dec-alg}} 
\vspace{-.15in}
\begin{center}
\fbox{%
\begin{minipage}{\textwidth}
\begin{small}
\alg{ERASURE-DECODE}($\mat{H}$,$\genericS{y}$)
\begin{algorithmic}[1]
\WHILE {$\genericS{y}$ \textnormal{has at least one erased sample}}
\IF {\textnormal{$\exists$ row $i$ of $\mat{H}$ (\ie, a check)
    connected to exactly one erased variable $y_j$}}
\STATE \textnormal{Set $y_j$ to be the XOR of all unerased bits in the
  check}
\ELSE
\STATE {\bf return} FAIL
\ENDIF
\ENDWHILE
\STATE \textnormal{Set $\genericS{x}$ to the message
  variables obtained from $\genericS{y}$}
\STATE {\bf return} \textnormal{the message variables $\genericS{x}$}
\end{algorithmic}
\end{small}
\end{minipage}
} 
\end{center}
\end{table}
\vspace{-.35in}

\subsubsection{Iterative Erasure Quantization}

The message-passing rules in \eqref{eq:bec-rules} can also be applied
to the BEQ problem for graphs without cycles provided some form of
tie-breaking is used.  Specifically, some variables will receive
erasure messages even after the algorithm has completed.  Such
variables can be arbitrarily chosen to be either 0 or 1 and still
produce a valid quantization.  For example in quantizing the source
$(*,*,1)$ with a (3,2) single parity check code, both $(1,0,1)$ and
$(0,1,1)$ are equally valid results and this tie can broken
arbitrarily.

On a graph with cycles, however, generalizing this approach by
initializing all unknown messages to $*$ usually fails.  For
example, on the dual of a Gallager code or a code like the one
represented in \figref{fig:beq_dual_ldpc} such an initialization rule
leads to all messages being erased at every step of the algorithm.
To perform effective tie-breaking, we need to distinguish between
variables which can be arbitrarily set to 0 or 1 and variables which
have not yet received enough information to be determined.  

One way to distinguish between these cases is to denote the former as
erasures with the symbol $*$ and the latter as null messages with the
symbol $\varnothing$, and initialize all messages to $\varnothing$.
With this initialization, we can use the following ``sum'' and
``product'' rules:
\footnote{In the product rule for erasure quantization, the symbol \#
denotes a contradiction.  If a contradiction is generated then
quantizing the given source with no distortion is impossible and the
algorithm fails.}

\begin{scriptsize}
\begin{equation}
\label{eq:beq-rules}
\begin{array}{c||c|c|c|c}
+ & 0 & 1 & * & \varnothing \\ \hline\hline
0 & 0 & 1 & * & \varnothing \\ \hline
1 & 1 & 0 & * & \varnothing \\ \hline
* & * & * & * & * \\ \hline
\varnothing & \varnothing & \varnothing & * & \varnothing 
\end{array}
\textnormal{\normalsize{ \ \ and \ \ }}
\begin{array}{c||c|c|c|c}
\times & 0 & 1 & * & \varnothing \\ \hline\hline
0      & 0 & \textnormal{\#}  & 0 & 0 \\ \hline
1      & \textnormal{\#} & 1 & 1 & 1 \\ \hline
*      & 0 & 1 & * & \varnothing \\ \hline
\varnothing      & 0 & 1 & \varnothing & \varnothing 
\end{array}
\end{equation}
\end{scriptsize}

\noindent Specifically, the outgoing message from a + node in a graph like
\figref{fig:beq_dual_ldpc} is computed by combining incoming messages
from all other edges with the $+$ rule.  The outgoing message from an
= node is computed by combining incoming messages from all other edges
with the $\times$ rule.  Whenever an = node has all incoming messages
being $*$, the value of the node is arbitrary.  This tie can
be broken by arbitrarily choosing a value of 0 or 1 provided the tie
is broken consistently.  Essentially, the requirement of consistent
tie-breaking can be interpreted as a constraint on the message-passing
schedule: tie-breaking information for a given tie should be propagated
through the graph before other ties are broken.

In order to provide a precise algorithm for the purpose of proving
theorems, we consider the \alg{ERASURE-QUANTIZE} in
\tabref{tab:erasure-quant-alg} based on applying the rules in
\eqref{eq:beq-rules} with a sequential schedule and all tie-breaking
collected into step~8.

\vspace{-.1in}
\begin{table}[htb]
\caption{An algorithm for iteratively quantizing a source with
  erasures. \label{tab:erasure-quant-alg}}
\vspace{-.15in}
\begin{center}
\fbox{%
\begin{minipage}{\textwidth}
\begin{small}
\alg{ERASURE-QUANTIZE}($\mat{G}$,$\genericS{z}$)
\begin{algorithmic}[1]
\WHILE {$\genericS{z}$ \textnormal{has at least one unerased sample}}
\IF {\textnormal{$\exists$ row $i$ of $\mat{G}$ (\ie, a variable)
    connected to exactly one unerased check $z_j$}}
\STATE \textnormal{Reserve message variable $i$ to later satisfy $z_j$ and
  erase check $z_j$}
\ELSE
\STATE {\bf return} FAIL \label{alg:e-quant-fail}
\ENDIF
\ENDWHILE
\STATE \textnormal{Arbitrarily set all unreserved message variables}
\STATE \textnormal{Set reserved variables to satisfy the corresponding
  checks starting
  from the last reserved variable and working backward to the first
  reserved variable} 
\STATE {\bf return} \textnormal{message variables $\genericS{w}$}
\end{algorithmic}
\end{small}
\end{minipage}
}
\end{center}
\end{table}
\vspace{-.35in}

\subsubsection{Iterative Algorithm Duality}

Our main results regarding iterative quantization are the following three
theorems stating that \alg{ERASURE-QUANTIZE} works correctly, can
be analyzed in the same manner as the \alg{ERASURE-DECODE} algorithm,
and works quickly:

\begin{thrm}
\label{thrm:eq-works}
For any linear code with generator matrix $\mat{G}$,
\alg{ERASURE-QUANTIZE}$(\mat{G},\genericS{z})$ either fails in
step~5 or else returns $\genericS{w}$ such that
$\genericS{w}\mat{G}$ matches $\genericS{z}$ in all unerased positions.
\end{thrm}

\begin{thrm}
\label{thrm:algs-fail-same}
Consider a linear code with parity check matrix $\mat{H}$ and its dual
code with generator matrix $\dualG = \mat{H}$.  The algorithm 
\alg{ERASURE-DECODE}$(\mat{H},\genericS{y})$ fails in step 5 if and only if
the algorithm \alg{ERASURE-QUANTIZE}$(\dualG,\genericS{z})$ fails in step
5 where $\genericS{y}$ has erasures specified by $\genericS{e}$ and $\genericS{z}$
has erasures specified by $\genericS{e}^{\perp} = 1 - \genericS{e}$.
\end{thrm}

\begin{thrm}
\label{thrm:e-quant-runs-quick}
The algorithm \alg{ERASURE-QUANTIZE}$(\mat{G},\genericS{z})$ runs in time
$\bigO(n\cdot d)$ where $n$ is the length of $\genericS{z}$ and $d$ is the
maximum degree of the graph corresponding to $\mat{G}$.
\end{thrm}

These results (proved in \appref{app:proofs}) imply that the parallel
structure between erasure decoding and erasure quantization allows us
to directly apply virtually every result from the analysis of one to
the other.  For example, these theorems combined with the
analysis/design of irregular LDPC codes achieving the capacity of the
binary erasure channel \cite{it:2002:oswald} immediately yield the
following Corollary:

\begin{corol}
There exists a sequence of linear codes which can be efficiently
encoded and decoded that achieves the rate-distortion function for binary
erasure quantization.
\end{corol}

\section{Concluding Remarks}
\label{sec:concluding-remarks}

In this paper we demonstrated how codes on sparse graphs combined with
iterative decoding can achieve the Shannon limit for binary erasure
quantization.  The main contribution of our algorithm is in
recognizing the role of tie-breaking, scheduling, and initialization
in iterative quantization.  The key insight in our analysis is the
strong dual relationship between error correction and quantization for
codes on graphs and their associated decoding/quantization algorithms
(both optimal and iterative).  We conjecture that the main task in
designing iterative message-passing algorithms for more general
quantization problems lies in designing appropriate tie-breaking,
scheduling, and initialization rules for such scenarios and exploiting
similar dual relationships to channel decoding.

\appendix
\section{Proofs}
\label{app:proofs}
\begin{proof}[Proof of \thrmref{th:ldpc-codes-bad}:]
Consider quantizing a random source and choose
some $c > 0$ and let $d$ be the smallest integer such that at
least $c \cdot n$ parity checks have degree at most $d$.  For each such
parity check, the probability that all variables in the check are not
erased is at least $(1-e)^d$.  Hence the probability that the check
cannot be satisfied is at least $(1/2) \cdot(1-e)^d$.  Since there are
$c \cdot n$ such checks, the probability that at least one check
cannot be satisfied is  
\begin{align}
\Pr[\textnormal{encoding failure}] &\geq 
1 - \left[1 - (1/2)\cdot\left(1-e\right)^d\right]^{c\cdot n} \\
&\geq
1 - \left[1 - \left(1/2-e/2\right)^d\right]^{c\cdot n} \\
&= 1 - \exp \left\{c\cdot n \ln \left[1 -
 \left(1/2-e/2\right)^d\right] \right\} \\
& \geq
1 - \exp \left\{-c\cdot n \cdot 
 \left(1/2-e/2\right)^d \right\} \\
&= 1 - \exp \left\{- \exp \left[ \ln c + \ln n + d \ln
 \left(1/2-e/2\right)\right] \right\}.
\end{align}
Hence for the probability of decoding failure to become small, $d$
must grow at least logarithmically with $n$ for every $c > 0$.  Note
that this argument applies to any parity-check graph representation of
the code.
\end{proof}

\begin{proof}[Proof of \thrmref{thrm:dual-optim-quant}:]
We will show that unique channel decoding is possible if and only if
the matrix equation $\mat{M} \genericS{x} = \genericS{y}$ has a solution (where
$\mat{M}$ will be defined shortly).  Similarly, we will show that
source quantization is possible for every $\genericS{z}$ if and only if the
matrix equation $\genericS{w} \mat{M} = \genericS{z}$ has a solution for every
$\genericS{z}$.  By demonstrating that both conditions are satisfied if and
only if the same matrix $\mat{M}$ has rank $n$, we will prove the
desired result.

Assume that all erasures occur in the last $\weight{\genericS{e}}$
positions (\ie, $\genericS{e} = 0^{n-\weight{\genericS{e}}}\
1^{\weight{\genericS{e}}} $).
\footnote{We use $\weight{\genericS{a}}$
to denote the number of non-zero values in $\genericS{a}$ (\ie, the
weight of $\genericS{a}$) and
$b^c \stackrel{\Delta}{=} \underbrace{(b\ b \ldots b)}_{\textnormal{c
    times}}$.}  
This incurs no loss of generality since the coordinates of $\cbook$
can always be permuted accordingly and the theorem applied to
the permuted code and its permuted dual code.
Let $\genericS{x}$ represent the transmitted signal and let $\genericS{y}$
denote the received signal.  Optimal decoding corresponds to finding a
vector which is a codeword of $\cbook$ and consistent with the
unerased received values.  The requirement that $\genericS{x}$ is a
codeword corresponds to the equation $\mat{H} \genericS{x} = 0$ where
$\mat{H}$ is the parity check matrix of $\cbook$.  The requirement
that $\genericS{x}$ is consistent with the received unerased data
corresponds to the equation $(\mat{I}_{n-\weight{\genericS{e}}}\ \mat{0}) \genericS{x} =
\vecIToJ{y}{1}{n-\weight{\genericS{e}}}$ where $\mat{I}_{t}$ represents a
$t$-by-$t$ identity matrix and $\vecIToJ{y}{i}{j}$ represents the
sub-vector $(y_i, y_{i+1}, \ldots, y_j)$.  Thus successful decoding is
possible if and only if the equation
\begin{equation}
\label{eq:bec-dec-cond}
\left(\begin{array}{cc}
\mat{I}_{n-\weight{\genericS{e}}} & \mat{0}\\
\multicolumn{2}{c}{\mat{H}}
\end{array}\right)
\genericS{x} = 
\left(\begin{array}{c}
\vecIToJ{y}{1}{n-\weight{\genericS{e}}}\\
0
      \end{array}
\right)
\end{equation}
has a unique solution.  According to well-known properties of linear
algebra, uniqueness is equivalent to the matrix in
\eqref{eq:bec-dec-cond} having full column rank (\ie, rank $n$).  Note
that existence of a solution is guaranteed since a
codeword was sent and no errors occurred.

Let $\genericS{z}$ represent the source to be quantized with erasure
pattern $\genericS{e}^{\perp}$.
Since we assumed that all erasures in $\genericS{e}$ occurred in the last
$\weight{\genericS{e}}$ samples, the dual erasure pattern
$\genericS{e}^{\perp}$ 
has all erasures occurring in the first $\weight{\genericS{e}^{\perp}}$
positions (\ie, $\genericS{e}^{\perp} = 
1^{n-\weight{\genericS{e}}}\ 0^{\weight{\genericS{e}}}$).
Optimal decoding corresponds to
finding a vector $\genericS{w}$ which is a codeword of $\cbook$ and
consistent with the unerased received values.  The former requirement
corresponds to the equation $\genericS{v}\dualG = \genericS{w}$ where
$\dualG$ is the generator matrix of $\dualcbook$ and
$\genericS{v}$ is a binary vector of appropriate dimension.  The
latter requirement corresponds to the equation
$\genericS{u}\left(\mat{I}_{\weight{\genericS{e}^{\perp}}}\ \mat{0}\right) +
\genericS{w} = \genericS{z}$ where $\genericS{u}$ is a binary vector chosen to ensure
that the first $\weight{\genericS{e}^{\perp}}$ positions (\ie, the erased positions)
match regardless of $\genericS{w}$.  Thus successful decoding is possible
for every $\genericS{z}$
if and only if a solution exists for 
\begin{equation}
\label{eq:beq-dec-cond}
\left(\begin{array}{cc}
\genericS{u} & \genericS{v}
      \end{array}
\right)
\left(\begin{array}{cc}
\mat{I}_{\weight{\genericS{e}^{\perp}}} & \mat{0}\\
\multicolumn{2}{c}{\dualG}
\end{array}\right)
= \genericS{z}
\end{equation}
for every $\genericS{z}$.
According to well-known properties of linear
algebra, existence of a solution for every $\genericS{z}$ is equivalent to
the matrix in 
\eqref{eq:beq-dec-cond} having full column rank (\ie, rank $n$).  Note
that uniqueness of a solution is neither guaranteed not required since
quantization is successful if at least one solution is found.

Noting that $\dualG = \mat{H}$ and $\weight{\genericS{e}^{\perp}}
= n-\weight{\genericS{e}}$ completes the proof since these conditions imply
that the matrices in \eqref{eq:bec-dec-cond} and
\eqref{eq:beq-dec-cond} are identical.
\end{proof}

\begin{proof}[Proof of \thrmref{thrm:eq-works}:]
For the algorithm to exit the while loop and reach step~8, every
unerased element of $z_i$ must have been erased in step~3 and assigned
a reserved message variable.  After a variable is reserved all its
checks must be erased.  Since checks can never changed from erased to
unerased, a reserved variable can never again be selected in step~2
and thus a variable can never be reserved more than once.

Thus after the while loop, each unerased position in $\genericS{z}$ has a
corresponding reserved variable.  Hence there exists an assignment of
the message variables which result in a codeword matching $\genericS{z}$ in
the unerased positions.  This assignment could be computed via
brute-force by solving he corresponding system of linear equations,
but in \thrmref{thrm:e-quant-runs-quick} we show that this 
step can be computed more efficiently.
\end{proof}

\begin{proof}[Proof of \thrmref{thrm:algs-fail-same}:]
The proof relies on the following invariant for steps~1--7 of both algorithms:
\begin{equation}
\label{eq:iter-inv}
\forall j, \ \textnormal{$y_j$ is erased if and only if $z_j$ is
unerased}.
\end{equation}
This condition is trivially true before the algorithm begins and forms
the base case for a proof by induction.  We assume that \eqref{eq:iter-inv}
holds at iteration $i$ of steps~1--7 and show that it must also hold
at iteration $i+1$.

First, \eqref{eq:iter-inv} implies that the outcome of step~1 is the
same for each algorithm.  Next, since $\dualG = \mat{H}$ the
tests in step~1 and step~2 of \alg{ERASURE-DECODE}$(\mat{H},\genericS{y})$
and \alg{ERASURE-QUANTIZE}$(\dualG,\genericS{z})$ yield the same
result.  Finally, at step~3, $y_j$ is unerased  while $z_j$ is
erased.  Therefore, by induction, condition \eqref{eq:iter-inv} is
true at every iteration and \alg{ERASURE-DECODE}$(\mat{H},\genericS{y})$
fails at step~5 if and only if
\alg{ERASURE-QUANTIZE}$(\dualG,\genericS{z})$ fails at step~5. 
\end{proof}

\begin{proof}[Proof of \thrmref{thrm:e-quant-runs-quick}:]
The while loop executes at most $n$ times.  Therefore step~1
requires at most $\bigO(n)$ operations.  Consider storing the number of
variables with exactly one unerased check in a data structure which
supports insertion and removal in constant time (\eg, a hash table).  
We can initialize the data structure with $\bigO(d\cdot n)$
operations.  Removing an element in steps~2 and 3 and updating the
data structure to account for step~3 requires $\bigO(d)$ operations.
Thus steps 1 through 8 require $\bigO(d\cdot n)$ operations and all
that remains is to bound the running time of step~9.

Denote the first reserved variable by $v_{j(1)}$, the second reserved
variable by $v_{j(2)}$ and so on to $v_{j(\weight{e})}$.  As
described in step~9, we first assign a value to
$v_{j(\weight{e})}$ and work backward.  Specifically, we set
$v_{j(i)}$ to the modulo-2 sum of $z_{j(i)}$ and all message variables
connected to $z_{j(i)}$ (except $v_{j(i)}$).  This is possible for
$z_{j(\weight{e})}$ since no other reserved variable could be
connected to $z_{j(\weight{e})}$.%
\footnote{If $z_{j(\weight{e})}$ was connected to another
  reserved variable $v$, that would imply $v$ was reserved when
  connected to $z_{j(\weight{e})}$ which was unerased as well as
  $z$ which must also have been unerased.  This contradicts step~2 in
 \alg{ERASURE-QUANTIZE}.}
Similarly, $z_{j(\weight{e}-1)}$ must be connected to only
unreserved variables as well as perhaps to
$v_{j(\weight{e})}$ and therefore a value can be determined
for $v_{j(\weight{e}-1)}$.  Thus, by induction we can
determine every $v_j$.

Adding up the operations computed for each step yields a running time
of $\bigO(d\cdot n)$.

\end{proof}

\paragraph{Acknowledgment}\ \\
The authors wish to thank G. D. Forney, Jr. for  many helpful comments
on the manuscript.

\begin{small}
\bibliographystyle{ieeetr}
\bibliography{paper}

\begin{thebibliography}{1}

\bibitem{Gray_1998}
R.~M. Gray and D.~L. Neuhoff, ``Quantization,'' {\em IEEE Transactions on
  Information Theory}, vol.~44, pp.~2325--2383, October 1998.

\bibitem{Marcellin_1990}
M.~W. Marcellin and T.~R. Fischer, ``Trellis coded quantization of memoryless
  and {G}auss-{M}arkov sources,'' {\em IEEE Transactions on Communications},
  vol.~38, pp.~82--93, January 1990.

\bibitem{it:2001:forney}
G.~D. Forney, Jr., ``Codes on graphs: normal realizations,'' {\em IEEE
  Transactions on Information Theory}, vol.~47, pp.~520--548, Feb 2001.

\bibitem{it:2002:oswald}
P.~Oswald and A.~Shokrollahi, ``Capacity-achieving sequences for the erasure
  channel,'' {\em IEEE Transactions on Information Theory}, vol.~48,
  pp.~3017--3028, December 2002.

\end{thebibliography}
\end{small}
\end{document}